\documentclass[prb,showpacs,twocolumn]{revtex4}%
\usepackage{amsfonts}
\usepackage{amsmath}
\usepackage{amssymb}
\usepackage{graphicx}%
\begin{document}
\title{Theory of conserved spin current and its application to two dimensional hole gas}
\author{Ping Zhang, Zhigang Wang}
\affiliation{Institute of Applied Physics and Computational Mathematics, P.O. Box 8009,
Beijing 100088, P.R. China}
\author{Junren Shi}
\affiliation{Institute of Physics, Chinese Academy of Sciences, Beijing 100080, P.R. China}
\author{Di Xiao, Qian Niu}
\affiliation{Department of Physics, The University of Texas at Austin, Austin, TX 78712}

\begin{abstract}
We present a detailed microscopic theory of the conserved spin current which
is introduced by us [Phys. Rev. Lett. \textbf{96}, 196602 (2006)] and
satisfies the spin continuity equation even for spin-orbit coupled systems.
The spin transport coefficients $\sigma_{\mu\nu}^{s}$ as a response to the
electric field are shown to consist of two parts, i.e., the conventional part
$\sigma_{\mu\nu}^{s0}$ and the spin torque dipole correction $\sigma_{\mu\nu
}^{s\tau}$. As one key result, an Onsager relation between $\sigma_{\mu\nu
}^{s}$ and other kinds of transport coefficients are shown. The expression for
$\sigma_{\mu\nu}^{s}$ in terms of single-particle Bloch states are derived, by
use of which we study the conserved spin Hall conductivity in the two
dimensional hole gas modeled by a combined Luttinger and SIA Rashba spin-orbit
coupling. It is shown that the two components in spin Hall conductivity
usually have the opposite contributions. While in the absence of Rashba spin
splitting, the spin Hall transport is dominated by the conventional
contribution, the presence of Rashba spin splitting stirs up a large
enhancement of the spin torque dipole correction, leading to an overall sign
change for the total spin Hall conductivity. Furthermore, an approximate
two-band calculation and the subsequent comparison with the exact four-band
results are given, which reveals that the coupling between the heavy hole and
light hole bands should be taken into account for strong Rashba spin splitting.

\end{abstract}
\maketitle

\section{Introduction}

Spintronics which combines the basic quantum mechanics of coherent spin
dynamics and technological applications in information processing and storage
devices\cite{Wolf,Awsch,Das}, has grown up to become a very active and
promising field in condensed matter. One central issue in spintronics is on
how to generate and manipulate spin current as well as to exploit its various
effects in a variety kinds of systems, ranging from ferromagnetic metals to
semiconductor paramagnets. In the ideal situation where spin is a good quantum
number, spin current is simply defined as the difference between the currents
of electron carried by the two spin states. This concept of the spin current
has served well in early studies of spin-dependent transport effects in
metals. The ubiquitous presence of spin-orbit coupling inevitably makes the
spin non-conserved, but this inconvenience is usually put off by focusing
one's attention within the so-called spin relaxation time. In recent years, it
has been found that the extrinsic or intrinsic spin-orbit coupling can provide
a route to generate transverse spin current in ferromagnetic
metals\cite{Hirsch,Zhang} or semiconductor paramagnets\cite{Dya,Muk1,Sinova}
by the driving of an electric field. The fundamental question of how to define
the spin current properly in the general situation then needs to be answered.
In most of previous studies of bulk spin transport, it has been conventional
to define the spin current simply as a combined thermodynamic and
quantum-mechanical average over the symmetric product of spin and velocity
operators. Unfortunately, no viable measurement is known to be possible for
this spin current. The recent spin-accumulation
experiments\cite{Kato,Wund,Sih} do not directly determine it, and there is no
deterministic relation between this spin current and the boundary spin accumulation.

In fact, the conventional definition of spin current suffers critical flaws
that prevent it from being relevant to spin transport and accumulation. First,
this spin current is not conserved, rendering it useless in describing a true
\textquotedblleft current\textquotedblright; Second, this spin current does
not necessarily vanish in insulators\cite{Fang,Mura2004}, and in thermodynamic
equilibrium\cite{Rashba2003}, so it is disqualified as a true transport
current corresponding to spin accumulation; Finally, there does not exist a
mechanical or thermodynamic force in conjugation with this current, so it
cannot be fitted into the standard near-equilibrium transport theory. The last
issue in particular makes the direct measurement of the conventional spin
current difficult if not impossible. For instance, because the conventional
spin transport coefficients cannot be associated with other transport
coefficients via Onsager relation\cite{PZhang}, they cannot be measured by
linking to other transport phenomena.

These issues were addressed in our last brief report\cite{Shi}, where we have
established a proper definition of spin current free from all the above
difficulties. The new spin current is given by the time derivative of the spin
displacement (product of spin and position observables), which differs from
the conventional definition by a torque dipole term. The torque dipole term is
first found in a semiclassical theory\cite{Culcer2004}, whose impact on spin
transport has been further analyzed to assess the importance of the inverse
spin Hall effect\cite{PZhang}, i.e., the charge Hall effect driven by a spin
force. The new spin transport coefficients $\sigma_{\mu\nu}^{s}$ are shown to
consist of two parts, i.e., the conventional part $\sigma_{\mu\nu}^{s0}$ and
the spin torque dipole correction $\sigma_{\mu\nu}^{s\tau}$. As one key
result, the Onsager relation between $\sigma_{\mu\nu}^{s}$ and other kinds of
force-driven transport coefficients has been shown. Note that the other
alternative definitions of spin current have also been proposed
recently\cite{Jin2005,Murakami2004,Zhang2005,Sun2005,Wang2006}.

In this paper, a detailed quantum-mechanical linear response theory of this
conserved spin current is addressed, which will be shown itself a necessary
supplement as well as an enlightening illustration for our previous brief
report\cite{Shi}. In particular, a general Kubo formula for the spin transport
coefficients $\sigma_{\mu\nu}^{s}$ in terms of single-particle Bloch states is
given in this paper. Then we use our formula to study the conserved spin Hall
conductivity in the two dimensional hole gas (2DHG) modeled by a combined bulk
Luttinger and space-inversion asymmetric (SIA) Rashba spin-orbit coupling. It
is shown that the two components in spin Hall conductivity usually have the
opposite contributions. While in the absence of Rashba spin splitting, the
spin Hall transport is dominated by the conventional contribution, the
presence of Rashba spin splitting stirs up a large enhancement of the spin
torque dipole correction, leading to an overall sign change for the total spin
Hall conductivity. Furthermore, an approximate two-band calculation and the
subsequent comparison with the exact four-band results are given, which
reveals that the coupling between the heavy hole and light hole bands should
be taken into account for strong Rashba spin splitting.

Our paper is organized as follows. In Sec. II we give a brief review of the
conserved spin current defined in Ref.\cite{Shi}. The intuitive arguments
begin with a search for spin density continuity equation by taking into
account the spin torque dipole correction. In Sec. III we clarify the
necessity to include the spin torque dipole correction into the total spin
current by initiating a linear response analysis. That is, only when this spin
torque dipole correction is included, can the proper Onsager relations between
the spin transport coefficients and the other transport coefficients be
established, which turn out to be essential to endow the spin current a
driving force. In Sec. IV we present a Kubo formula for the conserved spin
transport coefficients in terms of single-particle eigenstates. In Sec. V we
give a detailed discussion of the spin Hall effect in the two dimensional hole
gas by use of our general formulae. The analytic and numerical calculation are
given both in weak and strong SIA spin-orbit coupling regimes and show very
different features concerning the interplay between the two components in the
total spin Hall conductivity. Finally, in Sec. VI we present our conclusions.

\section{Spin continuity equation and introdution of conserved spin current}

The conventional definition of the spin current is the expectation value of
the product of spin and velocity operators. In the case of spin-polarized
magnetic systems, this spin current is simply reduced to the difference
between the spin-up and spin-down electrical currents. In the case of
spin-orbit stronly coupled semiconductor systems, where time-reversal symmetry
ensures no bulk spin polarization, unfortunately, no viable measurement is
known to be related with the conventional spin current. This can be
straightforwardly shown by writing down the continuity equation for spin
density,
\begin{equation}
\frac{\partial S_{z}}{\partial t}+\nabla\cdot\mathbf{J}_{s}=\mathcal{T}_{z}\,.
\tag{1}\label{E1}%
\end{equation}
The spin density for single-particle (spinor) state $\psi(\mathbf{r})$ is
defined by $S_{z}(\mathbf{r})=\psi^{\dagger}(\mathbf{r})\hat{s}_{z}%
\psi(\mathbf{r})$, where $\hat{s}_{z}$ is the spin operator for a particular
component ($z$ here, to be specific). The spin current density is given by the
conventional definition $\mathbf{J}_{s}(\mathbf{r})=\operatorname{Re}%
\psi^{\dagger}(\mathbf{r})\frac{1}{2}\{\hat{\mathbf{v}},\,\hat{s}_{z}%
\}\psi(\mathbf{r})$, where $\hat{\mathbf{v}}$ is the velocity operator, and
$\{,\}$ denotes the anticommutator. The right hand side of the Eq. (\ref{E1})
is the spin torque density defined by $\mathcal{T}_{z}(\mathbf{r}%
)=\operatorname{Re}\psi^{\dagger}(\mathbf{r})\hat{\tau}\psi(\mathbf{r})$,
where $\hat{\tau}\equiv d\hat{s}_{z}/dt\equiv(1/i\hbar)[\hat{s}_{z},\,\hat
{H}]$, and $\hat{H}$ is the Hamiltonian of the system. The presence of the
torque density $\mathcal{T}_{z}$ reflects the fact that spin is not conserved
microscopically in systems with spin-orbit coupling. As a consequence, the
only knowledge about the experimentally measured variation of spin density in
space-time is not sufficient to determine the conventional spin current
$\mathbf{J}_{s}$, and vice versa, due to the unique presence of the spin
torque density $\mathcal{T}_{z}$ in spin-orbit coupled systems.

One promising choice\cite{Shi} to remedy this oblique relationship between
spin density and spin current is to formaly move the torque density term to
the left hand side of Eq. (1) and absorb it in the divergence term. The
physical reason is that, due to symmetry reasons, it often happens that the
average torque vanishes for the bulk of the system, i.e., $(1/V)\int
dV\mathcal{T}_{z}(\mathbf{r})=0$. Thus one can write the torque density as a
divergence of a torque dipole density,
\begin{equation}
\mathcal{T}_{z}(\mathbf{r})=-\nabla\cdot\mathbf{P}_{\tau}(\mathbf{r})\,.
\tag{2}\label{E2}%
\end{equation}
Moving it to the left hand side of Eq. (\ref{E1}), one has
\begin{equation}
\frac{\partial S_{z}}{\partial t}+\nabla\cdot\left(  \mathbf{J}_{s}%
+\mathbf{P}_{\tau}\right)  =0\,, \tag{3}\label{E3}%
\end{equation}
which is in the form of the standard sourceless continuity equation. This
shows that the spin is conserved \textit{on average} in such systems, and the
corresponding transport spin current is:
\begin{equation}
\boldsymbol{\mathcal{J}}_{s}=\mathbf{J}_{s}+\mathbf{P}_{\tau}. \tag{4}%
\label{E4}%
\end{equation}
Note that there is still an arbitrariness in defining the effective spin
current because Eq.~(\ref{E2}) does not uniquely determine the torque dipole
density $\mathbf{P}_{\tau}$ from the corresponding torque density
$\mathcal{T}_{z}$. However, this ambiguity can be eliminated by imposing the
physical constraint that the torque dipole density is a material property that
should vanish outside the sample. This implies in particular that $\int
dV\,\mathbf{P}_{\tau}=-\int dV\,\mathbf{r}\nabla\cdot\mathbf{P}_{\tau}=\int
dV\,\mathbf{r}\mathcal{T}_{z}(\mathbf{r})$. It then follows that, upon bulk
average, the effective spin current density can be written as
$\boldsymbol{\mathcal{J}}_{s}=\operatorname{Re}\psi^{\ast}(\mathbf{r}%
)\hat{\boldsymbol{\mathcal{J}}}_{s}\psi(\mathbf{r})$, where
\begin{equation}
\hat{\boldsymbol{\mathcal{J}}}_{s}=\frac{\mathrm{d}(\hat{\mathbf{r}}\hat
{s}_{z})}{\mathrm{d}t}\, \tag{5}\label{E5}%
\end{equation}
is the effective spin current operator. Compared to the conventional spin
current operator, it has an extra term $\hat{\mathbf{r}}(\mathrm{d}\hat{s}%
_{z}/\mathrm{d}t)$, which accounts the contribution from the spin torque.

The conservation of the new spin current allows one to consider spin transport
in the bulk without the need of laboring explicitly a spin torque (dipole
density) which may be generated by the electric field. Thus the spin transport
in spin-orbit coupled systems can be treated in a unified way, whatever the
spin-orbit coupling strength is weak or strong. For example, it has been
customary to link spin density and spin current through the following
phenomenological spin continuity equation,%
\begin{equation}
\frac{\partial S_{z}}{\partial t}+\nabla\cdot\boldsymbol{\mathcal{J}}_{s}%
=-\frac{S_{z}}{\tau_{s}}, \tag{6}\label{E23}%
\end{equation}
where $\tau_{s}$ is the spin relaxation time, and the spin current has the
form $\boldsymbol{\mathcal{J}}_{s}=\sigma E-D_{s}\nabla S_{z}$. This makes
sense only if our new spin current is used in the calculation of spin
conductivity $\sigma$, otherwise an extra term of field-generated spin torque
must be added.

Equation (\ref{E23}) now can serve as the basis to determine the spin
accumulation at a sample boundary. Consider a system having a smooth boundary
produced by a slowly varying confining potential. We assume that the length
scale of variation is much larger than the mean free path, so that the above
continuity equation may be applied locally. By integrating from the interior
to the outside of the sample boundary, we obtain a spin accumulation per area
with $S_{z}=\boldsymbol{\mathcal{J}}_{s}^{\text{bulk}}\tau_{s}$. Thus one can
see that for the generic class of smooth boundaries, there is a unique
relationship between spin accumulation and the conserved spin current
$\boldsymbol{\mathcal{J}}_{s}$, instead of the conventional spin current
$\mathbf{J}_{s}$. The other kinds of boundary conditions have also been
discussed in previous work\cite{Shi,Nomura,Tse} to clarify the relationship
beween spin accumulation and bulk spin current.

\section{Spin Hall conductivity and its Onsager relation with inverse spin
Hall effect}

Defined as a time derivative of the spin displacement operatore $\hat
{\mathbf{r}}\hat{s}_{z}$, the new spin current has a natural conjugate force,
i.e., the spin force, $\mathbf{F}_{s}$. To show this, one can consider the
system exposed to an inhomogeneous Zeeman magnetic field along $z$-axis. The
resulting perturbation can be modeled as $V(\mathbf{r})=g^{\ast}\mu
_{B}B(\mathbf{r})\hat{s}_{z}$ with Bohr magneton $\mu_{B}$ and effective
magnetic factor $g^{\ast}$. Suppose that the inhomogeneous Zeeman field is
smoothly varying in space around zero. Then the first-order expansion in
position operator gives\cite{Zutic2002,PZhang}
\begin{equation}
V=-\mathbf{F}_{s}\cdot(\hat{\mathbf{r}}\hat{s}_{z}) \tag{7}\label{E6}%
\end{equation}
with a spin force $\mathbf{F}_{s}=-\nabla(g^{\ast}\mu_{B}B(\mathbf{r}))$
applying on the carriers. Thus it becomes clear now that only when the spin
current is defined as a time derivative of the spin displacement operator
$\hat{\mathbf{r}}\hat{s}_{z}$, can there naturally arises a conjugate driving
force $\mathbf{F}_{s}$. As a consequence, the energy dissipation rate for the
spin transport can be written as $\mathrm{d}Q/\mathrm{d}%
t=\boldsymbol{\mathcal{J}}_{s}{\small \cdot}\mathbf{F}_{s}$. It immediately
suggests a thermodynamic way to determine the spin current by simultaneously
measuring the Zeeman field gradient (spin force) and the heat generation.

The existence of a physical driving force for the conserved spin current makes
it possible to construct Onsager relation between spin transport coefficients
and other kinds of force-driving transport coefficients. For instance, since
an electric force $\mathbf{E}$ may drive a spin-Hall current through the
spin-orbit interaction, one naturally expects that a spin force $\mathbf{F}%
_{s}$ may also induce a charge-Hall current. Naturally, an exact Onsager
relation between these intrinsic Hall effects may be established: In a general
sense, when two kinds of different forces, say a spin force $\mathbf{F}_{s}$
and an electric force $\mathbf{E}$ coexist as driving forces, then the linear
charge-current and spin-current responses to these two forces can be expressed
as
\begin{equation}
\left(
\begin{array}
[c]{c}%
\boldsymbol{\mathcal{J}}_{s}\\
\mathbf{J}_{c}%
\end{array}
\right)  =\left(
\begin{array}
[c]{cc}%
\overleftrightarrow{\sigma}^{ss} & \overleftrightarrow{\sigma}^{sc}\\
\overleftrightarrow{\sigma}^{cs} & \overleftrightarrow{\sigma}^{cc}%
\end{array}
\right)  \left(
\begin{array}
[c]{c}%
\mathbf{F}_{s}\\
\mathbf{E}%
\end{array}
\right)  , \tag{8}\label{E7}%
\end{equation}
where $\boldsymbol{\mathcal{J}}_{s}$ is spin current and $\mathbf{J}_{c}$
charge current. $\overleftrightarrow{\sigma}^{ss}$ ($\overleftrightarrow
{\sigma}^{cc}$) is the spin-spin (charge-charge) $3\times3$ conductivity
tensor characterizing spin (charge) current response to a spin (charge) force
$\mathbf{F}_{s}$ ($\mathbf{E}$). In the same manner, the off diagonal block
$\overleftrightarrow{\sigma}^{sc}$ denotes spin current response to an
electric field, and $\overleftrightarrow{\sigma}^{cs}$ denotes charge current
response to a spin force\cite{PZhang}. A general relationship between
$\overleftrightarrow{\sigma}^{sc}$ and $\overleftrightarrow{\sigma}^{cs}$ can
be explicitly derived with a proper definition of the spin current, or imposed
by the general Onsager relation\cite{Casimir}:
\begin{equation}
\sigma_{\alpha\beta}^{sc}=\varepsilon_{\alpha}\varepsilon_{\beta}\sigma
_{\beta\alpha}^{cs} \tag{9}\label{E8}%
\end{equation}
where $\varepsilon_{\alpha}$, $\varepsilon_{\beta}$ are equal to $+1$ or $-1$
depending on whether the displacement (corresponding to current) operator is
even or odd under time reversal operation $T$. In the present case, the
displacement for a spin force is odd under $T$, while the displacement for an
electric field is $T$-invariant, implying $\varepsilon_{\alpha}\varepsilon
_{\beta}=-1$.

Here the key point is that the Onsager relation is only attainable when the
spin current is corrected by a torque dipole term. To see this more clearly,
we employ a standard Kubo formula description as follows: Let us put a spin
force along $\mu$-direction and an electric field $E$ along $\nu$-direction on
equal footing by including both of them in the total Hamiltonian,
$H=H_{0}-F_{1}d_{1}-F_{2}d_{2}$, where $F_{1}=E$ and $F_{2}=F^{s}$ are the
generalized forces applied on the charge and spin degrees of freedom, whereas
$d_{1}=-ex_{\mu}$ and $d_{2}=s_{z}x_{\nu}$ are the corresponding displacement
operators in which $x_{\mu}$ denotes the $\mu$-component of the position
observable $\mathbf{r}$. The response currents are obtained as the expectation
values of the generalized velocity operators in the perturbed states,
\begin{equation}
\dot{d}_{1}=-e\dot{x}_{\mu},\text{ }\dot{d}_{2}=s_{z}\dot{x}_{\nu}+\dot{s}%
_{z}x_{\nu}\mathbf{,}\tag{10}\label{E9}%
\end{equation}
where the standard symmetrization procedure in $\dot{d}_{2}$ is implied. The
presence of these two external forces will change an arbitrary stationary
quantum state $|\alpha\rangle$ of the original Hamiltonian into $|\alpha
^{\prime}\rangle=|\alpha\rangle+\sum_{\beta\neq\alpha,i=1,2}|\beta
\rangle\langle\beta|F_{i}d_{i}|\alpha\rangle/(\epsilon_{\alpha}-\epsilon
_{\beta}+i\eta)$, where $\epsilon_{\alpha}$ is unperturbed energy for
$|\alpha\rangle$ and $\eta\rightarrow0$ arises from the fact that when time
$t\rightarrow-\infty$, the perturbation is adiabatically switched off. Then
the currents as a linear response to the perturbations are, after a
straightforward derivation and by noting the substitution $\langle\beta
|d_{i}|\alpha\rangle=\frac{i\hbar}{\varepsilon_{\alpha}-\varepsilon_{\beta}%
}\langle\beta|\dot{d}_{i}|\alpha\rangle$, given by $\langle\dot{d}_{i}%
\rangle=\sum_{\alpha^{\prime}}f_{\alpha^{\prime}}\langle\alpha^{\prime}%
|\dot{d}_{i}|\alpha^{\prime}\rangle\equiv\sum_{j}\sigma_{ij}F_{j}$ with the
conductivity matrix
\begin{align}
\sigma_{ij} &  =\sum_{\alpha\neq\beta}\frac{\hbar\operatorname{Im}%
[\langle\alpha|\dot{d}_{i}|\beta\rangle\langle\beta|\dot{d}_{j}|\alpha
\rangle]}{(\epsilon_{\alpha}-\epsilon_{\beta})^{2}+\eta^{2}}(f_{\alpha
}-f_{\beta})\tag{11}\label{E10}\\
&  +\sum_{\alpha\neq\beta}\frac{df_{\alpha}(\epsilon)}{d\epsilon
}\operatorname{Re}[\langle\alpha|\dot{d}_{i}|\beta\rangle\langle\beta|\dot
{d}_{j}|\alpha\rangle]\delta(\epsilon_{\alpha}-\epsilon_{\beta})\nonumber
\end{align}
Here $f_{\alpha}$ is the equilibrium fermi function for band $\epsilon
_{\alpha}$. Obviously, the first term is antisymmetric under the
inter-exchange $i\leftrightarrow j$, while the second term is the symmetric
part of $\sigma_{ij}$. While the symmetric part denotes dissipative
contribution to the charge or spin transport, the first antisymmetric term is
dissipationless and is relevant to our discussion in this paper. For an ideal
system without scattering and many-body interaction, it is obvious that the
dissipative (symmetric) part in Eq.(\ref{E10}) vanishes. The dissipationless
conductivity has manifested itself in a fundamental way in condensed matter
physics involving such issues as quantum and anomalous Hall effects. Our
present discussions, as well as recent active studies of spin-Hall effect, are
also focused on this dissipationless part of the transport conductivity.

It becomes clear from Eqs.(\ref{E9})-(\ref{E10}) that the intrinsic spin-Hall
conductivity is given by
\begin{equation}
\sigma_{\mu\nu}^{s}=-e\sum_{\alpha\neq\beta}\frac{\hbar\operatorname{Im}%
[\langle\alpha|s_{z}\dot{x}_{\mu}+\dot{s}_{z}x_{\mu}|\beta\rangle\langle
\beta|\dot{x}_{\nu}|\alpha\rangle]}{(\varepsilon_{\alpha}-\varepsilon_{\beta
})^{2}+\eta^{2}}(f_{\alpha}-f_{\beta}), \tag{12}\label{E11}%
\end{equation}
where again symmetrization is implied in the product of two operators. On the
other side, the dissipationless inverse spin-Hall conductivity driven by a
spin force is given by%
\begin{equation}
\sigma_{\nu\mu}^{c}=-e\sum_{\alpha\neq\beta}\frac{\hbar\operatorname{Im}%
[\langle\alpha|\dot{x}_{\nu}|\beta\rangle\langle\beta|s_{z}\dot{x}_{\mu}%
+\dot{s}_{z}x_{\mu}|\alpha\rangle]}{(\varepsilon_{\alpha}-\varepsilon_{\beta
})^{2}+\eta^{2}}(f_{\alpha}-f_{\beta}) \tag{13}\label{E12}%
\end{equation}
The antisymmetric property shown above immediately gives the equality%
\begin{equation}
\sigma_{\mu\nu}^{s}=-\sigma_{\nu\mu}^{c} \tag{14}\label{E13}%
\end{equation}
which coincides with the general Onsager relation (\ref{E8}). Note that
Eq.(\ref{E13}) denotes the Onsager relation only for the dissipationless
(antisymmetric) part in conductivities. A full Onsager relation by taking into
account impurity scattering and an external uniform magnetic field $B_{0}$
turns out to be given by $\sigma_{\mu\nu}^{s}(B_{0},\tau_{0})=-\sigma_{\nu\mu
}^{c}(-B_{0},-\tau_{0})$, where $\tau_{0}$ characterizes scattering-induced
relaxation time. Therefore, from Eqs.(\ref{E11})-(\ref{E12}) one can see that
to ensure the Onsager relation between the two kinds of Hall conductivities,
the spin current should be defined as
\begin{equation}
\boldsymbol{\mathcal{J}}_{s}=\frac{\mathrm{d}(\hat{\mathbf{r}}\hat{s}_{z}%
)}{\mathrm{d}t} \tag{15}\label{E14}%
\end{equation}
instead of conventional definition.

The above arguments clearly show two prominent features of the new spin
current which is absent within the conventional definition of the spin
current: (i) The spin current is now conserved with a physical conjugate
driving force; (ii) The Onsager relation is built up. Besides these two
prominent featues, there is another physically valid property: For simple
insulators whose single particle eigenstates are localized (Anderson
insulators), the spin transport coefficients vanish. Indeed, for spatially
localized eigenstates, we can evaluate the (intrinsic) conductivity from Eq.
(\ref{E11}) as,%
\begin{align}
\sigma^{s}  &  =-e\hbar\sum_{\alpha\neq\beta}f_{\alpha}\frac{\operatorname{Im}%
[\langle\alpha|\mathrm{d}(\hat{\mathbf{r}}\hat{s}_{z})/\mathrm{d}%
t|\beta\rangle\langle\beta|\mathbf{\hat{v}}|\alpha\rangle]}{(\epsilon_{\alpha
}-\epsilon_{\beta})^{2}}\tag{16}\label{E15}\\
&  =-e\hbar\sum_{\alpha}f_{\alpha}\langle\alpha|[\hat{\mathbf{r}}\hat{s}%
_{z},\hat{\mathbf{r}}]|\alpha\rangle=0\nonumber
\end{align}
where we have used $\langle\alpha|\mathrm{d}(\hat{\mathbf{r}}\hat{s}%
_{z})/\mathrm{d}t|\beta\rangle=\frac{i}{\hbar}(\epsilon_{\alpha}%
-\epsilon_{\beta})\langle\alpha|\hat{\mathbf{r}}\hat{s}_{z}|\beta\rangle$ and
$\langle\beta|\mathbf{\hat{v}}|\alpha\rangle=\frac{i}{\hbar}(\epsilon_{\beta
}-\epsilon_{\alpha})\langle\beta|\hat{\mathbf{r}}|\alpha\rangle$. The involved
matrix elements are well defined between spatially localized eigenstates.

\section{Kubo formula for conserved spin transport coefficients}

In this section we show how to evaluate in practice the conserved spin Hall
conductivity based on the new definition of spin current in a crystal. A
formal description has been given in Eq. (10) for the conserved spin
conductivity, while the general states $|\alpha\rangle$ should now be replaced
by the electron Bloch wave function $\psi_{nk}(\mathbf{r})$. Clearly, the
conserved spin Hall conductivity includes two components,
\begin{equation}
\sigma_{\mu\nu}^{s}=\sigma_{\mu\nu}^{s0}+\sigma_{\mu\nu}^{s\tau}.
\tag{17}\label{E16}%
\end{equation}
The first one is the usual conventional spin Hall coefficient, which is ready
to be rewritten in the Bloch-state space as
\begin{align}
\sigma_{\mu\nu}^{s0}  &  =-e\hbar\sum_{n\neq n^{\prime},\mathbf{k}}\left[
f(\epsilon_{n\mathbf{k}})-f(\epsilon_{n^{\prime}\mathbf{k}})\right]
\times\tag{18}\label{E17}\\
&  \frac{\operatorname{Im}\langle u_{n\mathbf{k}}|\frac{1}{2}\{\hat{v}_{\mu
},\hat{s}_{z}\}|u_{n^{\prime}\mathbf{k}}\rangle\langle u_{n^{\prime}%
\mathbf{k}}|\hat{v}_{\nu}|u_{n\mathbf{k}}\rangle}{\left(  \epsilon
_{n\mathbf{k}}-\epsilon_{n^{\prime}\mathbf{k}}\right)  ^{2}+\eta^{2}%
}.\nonumber
\end{align}
Here $\mathbf{\hat{v}}(\mathbf{k})=\partial H(\mathbf{k})/\hbar\partial
\mathbf{k}$, $H(\mathbf{k})=\exp(-i\mathbf{k}{\small \cdot}\mathbf{r})\hat
{H}\exp(i\mathbf{k}{\small \cdot}\mathbf{r})$, $|u_{n\mathbf{k}}\rangle$ is
the periodic part of the electron carrier Bloch wave function, and ($-e$) is
the electron charge. The Kubo formula (\ref{E17}) for the conventional spin
Hall conductivity has been used by most of previous
investigations\cite{Sinova}.

The second component $\sigma_{\mu\nu}^{s\tau}$ in the conserved spin-Hall
conductivity, as shown in Eq. (\ref{E4}) and Eq. (\ref{E11}), comes from the
contribution of the spin torque density term. Due to the\ presence of the
position operator in this term, and the fact that the position operator is not
well-defined in the Bloch-state space, one needs to take some care in dealing
with $\sigma_{\mu\nu}^{s\tau}$. As one choice for derivation, we proceed by a
regularization scheme, $\dot{s}_{z}x_{\mu}\rightarrow\dot{s}_{z}\sin(q_{\mu
}x_{\mu})/q_{\mu}$, followed by the limit $q_{\mu}\rightarrow0$. The
symmetrized $\mu$-component of the spin torque density operator occurred in
Eq. (\ref{E11}) can now be rewritten as
\begin{equation}
P_{\tau}=\frac{1}{2}\{\dot{s}_{z},\frac{\sin(qx_{\mu})}{q}\}=\lim
_{q\rightarrow0}\frac{1}{4iq}\left[  \{\dot{s}_{z},e^{iqx_{\mu}}\}-\{\dot
{s}_{z},e^{-iqx_{\mu}}\}\right]  \tag{19}\label{E18}%
\end{equation}
Substituting Eq. (\ref{E18}) into Eq. (\ref{E11}) and after a straightforward
manipulation, we obtain the Kubo formula for $\sigma_{\mu\nu}^{s\tau}$:%
\begin{align}
\sigma_{\mu\nu}^{s\tau}  &  =-e\hbar\lim_{\mathbf{q}\rightarrow0}\frac
{1}{q_{\nu}}\sum_{n\neq n^{\prime},\mathbf{k}}\left[  f(\epsilon_{n\mathbf{k}%
})-f(\epsilon_{n^{\prime}\mathbf{k+q}})\right]  \times\tag{20}\label{E19}\\
&  \frac{\operatorname{Re}[\langle u_{n\mathbf{k}}|\hat{\tau}(\mathbf{k}%
,\mathbf{q})|u_{n^{\prime}\mathbf{k}+\mathbf{q}}\rangle\langle u_{n^{\prime
}\mathbf{k}+\mathbf{q}}|\mathbf{\hat{v}}(\mathbf{k},\mathbf{q})|u_{n\mathbf{k}%
}\rangle]}{\left(  \epsilon_{n\mathbf{k}}-\epsilon_{n^{\prime}\mathbf{k+q}%
}\right)  ^{2}+\eta^{2}},\nonumber
\end{align}
where $\hat{\tau}(\mathbf{k},\mathbf{q})=\frac{1}{2}\left[  \hat{\tau
}(\mathbf{k})+\hat{\tau}(\mathbf{k}+\mathbf{q})\right]  $ with $\hat{\tau
}(\mathbf{k})=(1/i\hbar)[\hat{s}_{z},H(\mathbf{k})]$, $\mathbf{\hat{v}%
}(\mathbf{k},\mathbf{q})=\frac{1}{2}[\mathbf{\hat{v}}(\mathbf{k}%
)+\mathbf{\hat{v}}(\mathbf{k}+\mathbf{q})]$. The limit of $\eta\rightarrow0$
should be taken at the last step of calculation, and as a result, there is no
intra-band ($n=n^{\prime}$) contribution. Note that to properly calculate
$\sigma_{\mu\nu}^{s\tau}$ in practice, all the terms in Eq. (4) with the
subscript $\mathbf{k}+\mathbf{q}$ should be expaned at $\mathbf{k}$ to first
order in $\mathbf{q}$.

Equation (\ref{E19}) can also be derived from an analysis of the spin torque
dipole density $\mathcal{T}_{z}(\mathbf{r})$ which can be determined
unambiguously as a bulk property within the theoretical framework of linear
response. Consider the torque response to an electric field at finite wave
vector $\mathbf{q}$, $\mathcal{T}_{z}(\mathbf{q})=\mathbf{\chi}(\mathbf{q}%
){\small \cdot}\mathbf{E(q)}$. Based on Eq. (\ref{E2}) which implies
$\mathcal{T}_{z}(\mathbf{q})=-i\mathbf{q}{\small \cdot}\mathbf{P}_{\tau
}\mathbf{(q)}$, we can uniquely determine the static response (i.e.,
$\mathbf{q}\rightarrow0$) of the spin torque dipole:%
\begin{equation}
\mathbf{P}_{\tau}\mathbf{(q)=\operatorname{Re}}\{i\nabla_{\mathbf{q}%
}[\mathbf{\chi}(\mathbf{q}){\small \cdot}\mathbf{E}]\}_{\mathbf{q}=0}.
\tag{21}\label{E20}%
\end{equation}
Here we have utilized the condition $\mathbf{\chi}(0)=0$, i.e., there is no
bulk spin generation by the electric field. Thus the consequent spin-transport
coefficient for the electric response of the spin torque density is given by
\begin{equation}
\sigma_{\mu\nu}^{s\tau}=\operatorname{Re}[i\partial_{q_{\mu}}\chi_{\nu
}(\mathbf{q})]_{\mathbf{q}=0}. \tag{22}\label{E21}%
\end{equation}
Again within the standard Kubo-formula formalism, the spin torque response
coefficient $\mathbf{\chi}(\mathbf{q})$ can be calculated to be%
\begin{align}
\mathbf{\chi}(\mathbf{q})  &  =ie\hbar\sum_{n\neq n^{\prime},\mathbf{k}%
}\left[  f(\epsilon_{n\mathbf{k}})-f(\epsilon_{n^{\prime}\mathbf{k+q}%
})\right]  \times\tag{23}\label{E22}\\
&  \frac{\operatorname{Re}[\langle u_{n\mathbf{k}}|\hat{\tau}(\mathbf{k}%
,\mathbf{q})|u_{n^{\prime}\mathbf{k}+\mathbf{q}}\rangle\langle u_{n^{\prime
}\mathbf{k}+\mathbf{q}}|\mathbf{\hat{v}}(\mathbf{k},\mathbf{q})|u_{n\mathbf{k}%
}\rangle]}{\left(  \epsilon_{n\mathbf{k}}-\epsilon_{n^{\prime}\mathbf{k+q}%
}\right)  ^{2}+\eta^{2}},\nonumber
\end{align}
which, combining Eq. (\ref{E21}), gives the same expression for $\sigma
_{\mu\nu}^{s\tau}$ as that in Eq. (\ref{E19}).

Due to the inclusion of contribution from the spin torque dipole term, one may
expect that in some special cases the transport properties of the conserved
spin current are essentially different from that of conventional spin current.
This will turn to be true by studying in the next section the intrinsic spin
Hall conductivity in a 2DHG system, which is modeled by a combined Luttinger
and spin-3/2 SIA Rashba spin-orbit interactions. For disordered systems, the
spin-Hall conductivity based on our new spin current has also been calculated
in a recent work\cite{Sugimoto}, and is found to depend explicitly on the
scattering potentials for the two dimensional Rashba models with $k$-linear or
$k$-cubic spin-orbit coupling. For one who instead prefers the conventional
definition of the spin current, the above discussions are of course still
helpful due to the fact that one cannot at last avoid tackling the calculation
of the spin torque contribution to the total transport spin current and the
spin accumulation, while our above derivation clearly indicates how to do that
in practice. \ 

\section{Application to 2DHG system}

As a practical application of the above general theory of the conserved spin
current to the real physical systems, in this section we focus our attention
to the intrinsic spin Hall effect in a 2DHG system, which has been
experimentally\cite{Wund} and theoretically\cite{Bern} investigated within the
conventional spin-transport framework. Following Ref.\cite{Bern}, the
spin-orbit interaction in this system is modeled by combined Luttinger and
spin-3/2 SIA Rashba terms. The resultant $\mathbf{k}{\small \cdot}\mathbf{p}%
$\ Hamiltonian reads ($\hbar$ is set to be unity):
\begin{equation}
H_{0}=(\gamma_{1}+\frac{5}{2}\gamma_{2})\frac{k^{2}}{2m}-\frac{\gamma_{2}}%
{m}(\mathbf{k{\small \cdot}S})^{2}+\alpha(\mathbf{S}\times\mathbf{k}%
){\small \cdot}\hat{z}, \tag{24}\label{E24}%
\end{equation}
where the confinement of the quantum well in the $\hat{z}$ direction makes the
momentum be quantized on this axis. The crucial difference between SIA Rashba
term in 2DHG and SIA Rashba term in the 2DEG lies in the fact that $S$ in 2DHG
is spin-3/2 matrix, describing both the heavy (HH) and light (LH) holes. For
the first heavy and light hole bands, the confinement in a quantum well of
thickness $a$ is approximated by the relation $\langle k_{z}\rangle=0$,
$\langle k_{z}^{2}\rangle\approx(\pi/a)^{2}$. The energies for HH and LH are
given by \begin{widetext}
\begin{align*}
E_{\pm}^{LH} &  =\frac{\gamma_{1}}{2m}(k^{2}+\langle
k_{z}^{2}\rangle)\pm
\frac{1}{2}\alpha k-\sqrt{\alpha^{2}k^{2}\pm\frac{\alpha\gamma_{2}}{m}%
k(k^{2}+\langle k_{z}^{2}\rangle)+\frac{\gamma_{2}^{2}}{m^{2}}d^{2}},\tag{25}\label{E25}\\
E_{\pm}^{HH} &  =\frac{\gamma_{1}}{2m}(k^{2}+\langle
k_{z}^{2}\rangle)\pm
\frac{1}{2}\alpha k+\sqrt{\alpha^{2}k^{2}\pm\frac{\alpha\gamma_{2}}{m}%
k(k^{2}+\langle k_{z}^{2}\rangle)+\frac{\gamma_{2}^{2}}{m^{2}}d^{2}},%
\end{align*}
\end{widetext}where and in the following $k^{2}=k_{x}^{2}+k_{y}^{2}$,
$d=\sqrt{k^{4}+\langle k_{z}^{2}\rangle^{2}-k^{2}\langle k_{z}^{2}\rangle}$.
The HH and LH bands are split at the $\Gamma$ point by $\Delta=2\gamma
_{2}\langle k_{z}^{2}\rangle/m$. Depending on the confinement scale $a$ the
Luttinger term is dominant for $a$ not too small, while the SIA term becomes
dominant for thin quantum wells.

By expanding the above formulas for small $k<<\langle k_{z}\rangle$ it is seen
that the spin splitting of the HH bands is $k^{3}$ whereas the spin splitting
of the LH bands is $k$\cite{Bern},
\begin{align}
E_{+}^{HH}-E_{-}^{HH}  &  =\frac{3}{8}\frac{\alpha(\alpha^{2}-4\frac
{\gamma_{2}^{2}}{m^{2}}\langle k_{z}^{2}\rangle)}{\frac{\gamma_{2}^{2}}{m^{2}%
}\langle k_{z}^{2}\rangle^{2}}k^{3}+\mathcal{O}(k^{5})\tag{26}\label{E26}\\
E_{+}^{LH}-E_{-}^{LH}  &  =2\alpha k+\mathcal{O}(k^{3})\nonumber
\end{align}
which is in agreement with\cite{Wink1,Wink2}. Figure 1(a) gives a typical band
structure for GaAs ($\gamma_{1}=6.92$, $\gamma_{2}=2.1$) with a $\Gamma$ point
gap of $40$ meV and a Fermi momentum splitting of the hole band at Fermu
momentum ($0.2$ nm$^{-1}$) of 5 meV, which requires a SIA splitting
$\alpha\approx50$ meV${\small \cdot}$nm. In the recent experiment of spin-Hall
effect\cite{Wund}, this energy gap is of order $\triangle E=40$meV, which
corresponds to an $a=8.3$ nm thick quantum well.%
\begin{figure}[tbp]
\begin{center}
\includegraphics[width=1.0\linewidth]{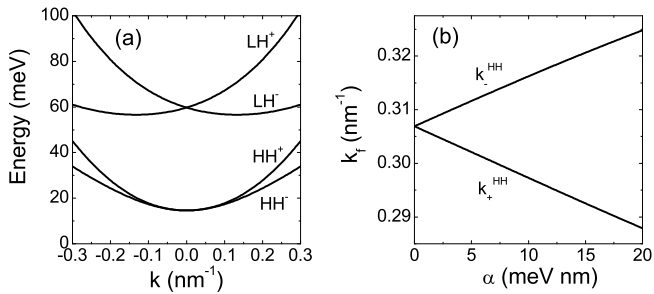}
\end{center}
\caption
{(a) Approximate band structure of the GaAs 2DHG; (b) heavy-hole fermi wave vectors as a function Rashba coefficient $\alpha
$ for $n_{h}=1.5\times10^{12}$ cm$^{-2}$.} \label{fig1}
\end{figure}
During the following numerical calculation, these three material parameters
($\gamma_{1}$, $\gamma_{2}$, and $a$) will be fixed to be the values mentioned
above, whereas the Rashba coefficient $\alpha$ and the hole density $n_{h}$
are treated as the tuning parameters.

First we consider the simple case of thin quantum wells. In this case the SIA
Rashba term can be neglected ($\alpha=0$) and a full analytic calculation an
be carried out.\ In the absence of SIA term, the (normalized) eigenstates for
HH and LH bands are given by $u_{+}^{HH}=a_{1}|\frac{3}{2},\frac{3}{2}%
\rangle+a_{2}k_{+}^{2}|\frac{3}{2},-\frac{1}{2}\rangle$, $u_{-}^{HH}%
=a_{1}|\frac{3}{2},-\frac{3}{2}\rangle+a_{2}k_{-}^{2}|\frac{3}{2},\frac{1}%
{2}\rangle$, $u_{+}^{LH}=a_{1}|\frac{3}{2},\frac{1}{2}\rangle-a_{2}k_{+}%
^{2}|\frac{3}{2},-\frac{3}{2}\rangle$, and $u_{-}^{LH}=a_{1}|\frac{3}%
{2},-\frac{1}{2}\rangle-a_{2}k_{-}^{2}|\frac{3}{2},\frac{3}{2}\rangle$, where
$|\frac{3}{2},m_{z}\rangle$ ($m_{z}=\frac{3}{2}$,...$-\frac{3}{2}$) are the
eigenstates of $S_{z}$, $k_{\pm}=k_{x}\pm ik_{y}$, $a_{1}=\sqrt{(d-\frac{1}%
{2}k^{2}+\langle k_{z}^{2}\rangle)/2d}$, $a_{2}=\sqrt{3/8d(d-\frac{1}{2}%
k^{2}+\langle k_{z}^{2}\rangle)}$. Then using the standard Kubo formula
(\ref{E18}) for the conventional spin Hall conductivity, it is straightforward
to obtain $\sigma_{xy}^{s0}$ as follows \begin{widetext}
\[
\sigma_{yx}^{0}=-\frac{e}{4\pi}\int(f_{HH}-f_{LH})\left[  \frac{3k^{2}}{d^{2}%
}(a_{1}^{4}-a_{2}^{4}k^{8})+\frac{2\sqrt{3}k^{2}}{\gamma_{2}d^{2}}a_{1}%
a_{2}k^{2}(a_{1}^{2}+a_{2}^{2}k^{4})(\gamma_{1}+\gamma_{2})\right]
kdk \tag{27} \label{E27}.
\]
\end{widetext}where $f_{HH}$ and $f_{LH}$ are fermi functions for HH and LH
respectivel. The spin Hall conductivity due to the contribution from the
torque dipole term turns out to be \begin{widetext}
\begin{align*}
\sigma_{yx}^{s\tau}  & =\frac{e}{4\pi}\frac{\gamma_{1}}{\gamma_{2}}\int
(f_{HH}-f_{LH})\left[  \frac{2\sqrt{3}a_{1}a_{2}k^{4}}{d^{2}}(a_{1}^{2}%
+a_{2}^{2}k^{4})+\frac{3k^{4}}{d^{3}}(a_{1}^{2}+a_{2}^{2}k^{4})^{2}\right]
kdk \tag{28} \label{E28}\\
& +\frac{e}{4\pi}\int\frac{3k^{4}(a_{1}^{2}+a_{2}^{2}k^{4})^{2}}{2d^{2}%
}\left(  \frac{df_{HH}}{dk}+\frac{df_{LH}}{dk}\right)  dk
\end{align*}
\end{widetext}The first line in Eq. (\ref{E28}) is obtained by expanding
$|u_{n^{\prime}\mathbf{k}+\mathbf{q}}\rangle$, $\hat{\tau}(\mathbf{k}%
,\mathbf{q})$, and $\epsilon_{n^{\prime}\mathbf{k}+\mathbf{q}}$ in Eq.
(\ref{E19}) at $\mathbf{k}$ to the first order in $\mathbf{q}$ while remaining
other quantities to be their values at $\mathbf{q}=0$. Whereas the second line
in Eq. (\ref{E28}) with $df/dk$ is obtained by a linear expansion of the fermi
function with respect to $\mathbf{q}$. The other terms occurring in Eq.
(\ref{E19}) turn out to take no contribution to $\sigma_{yx}^{s\tau}$ for the
present model. In the strong confinement limit ($k^{2}<\langle k_{z}%
^{2}\rangle$) and at zero temperature, Eqs. (\ref{E27})-(\ref{E28}) are
reduced to
\begin{align}
\sigma_{yx}^{s0} &  =-\frac{e}{4\pi}\left[  \frac{3(k_{HH}^{4}-k_{LH}^{4}%
)}{4\langle k_{z}^{2}\rangle^{2}}+\frac{(\gamma_{1}+\gamma_{2})(k_{HH}%
^{6}-k_{LH}^{6})}{4\gamma_{2}\langle k_{z}^{2}\rangle^{3}}\right]
,\nonumber\\
\sigma_{yx}^{s\tau} &  =\frac{e}{4\pi}\left[  \frac{3\gamma_{1}(k_{HH}%
^{6}-k_{LH}^{6})}{4\gamma_{2}\langle k_{z}^{2}\rangle^{3}}-\frac{3(k_{HH}%
^{4}+k_{LH}^{4})}{2\langle k_{z}^{2}\rangle^{2}}\right]  .\tag{29}\label{a1}%
\end{align}
where $k_{HH}$ and $k_{LH}$ are the fermi wavevectors for HH and LH respectively.%

\begin{figure}[tbp]
\begin{center}
\includegraphics[width=1.0\linewidth]{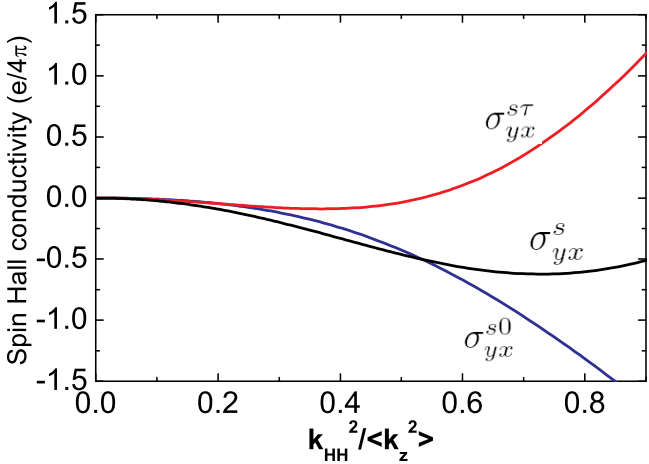}
\end{center}
\caption{(Color online) conserved spin Hall conductivity $\sigma_{yx}%
^{s}$ (black line), and its two
components $\sigma_{yx}^{s0}$ (blue line) and $\sigma_{yx}^{s\tau}$ (red line)
as a function of the heavy hole fermi wave vector for $\alpha=0$ and $n_{h}%
=1.5\times10^{12}$ cm$^{-2}$.} \label{fig2}
\end{figure}%
Since for the experiment data, the LH bands are unoccupied (by the holes), so
during numerical calculations throughout this paper we have set the fermi
wavevector of the light holes to be zero, $k_{LH}=0$. Based on Eqs.
(\ref{E27})-(\ref{E28}), figure 2 shows the conserved spin Hall conductivity
$\sigma_{yx}^{s}$ (black curve) as a function of the fermi wavevector $k_{HH}$
(scaled by the confinement wave vector $\sqrt{\langle k_{z}^{2}\rangle}$) of
the heavy holes. For clarification, we have also plotted in Fig. 2 the
separate contributions from the conventional part $\sigma_{yx}^{s0}$ (blue
curve) and spin torque dipole part $\sigma_{yx}^{s\tau}$ (red cuve); their sum
gives $\sigma_{yx}^{s}$. With increasing the fermi wavevector $k_{HH}$, one
can see from Fig. 2 that the two components $\sigma_{yx}^{s0}$ and
$\sigma_{yx}^{s\tau}$ increase in amplitude but always differ in a sign. The
resultant total spin Hall conductivity $\sigma_{yx}^{s}$ displays a
non-monotonic behavior. The typical experiments\cite{Wund} are usually carried
out in the region of small value of fermi wavevector $k_{HH}$ (compared to
$\sqrt{\langle k_{z}^{2}\rangle}$). In this region, it reveals in Fig. 2 that
the tendency of the total SHC aligns with that of the conventional one with
very little difference. Under experimental condition $\sqrt{\langle k_{z}%
^{2}\rangle}=0.38$ nm$^{-1}$ and $k_{HH}=0.31$, the conventional term is
obtained to $\sigma_{yx}^{s0}=-\frac{0.8e}{4\pi}$ (compared to the
previous\cite{Bern} theoretical calculation result of $-\frac{0.9e}{4\pi}$),
while the spin torque dipole term is given by $\sigma_{yx}^{s\tau}=\frac
{0.2e}{4\pi}$. The total spin conductance is therefore $\sigma_{yx}^{s}%
=-\frac{0.6e}{4\pi}$, with a tiny deviation from the value of conventional one
$\sigma_{yx}^{s0}$. We notice that the numerical simulation\cite{Wund} related
to the experimental setup gives a similar value of $\sigma_{yx}^{s0}$,
although the experiment was done in strong SIA Rashba spin-orbit coupling
regime, while the present calculation with the result given in Fig. 2 is for
$\alpha=0$.%

\begin{figure}[tbp]
\begin{center}
\includegraphics[width=1.0\linewidth]{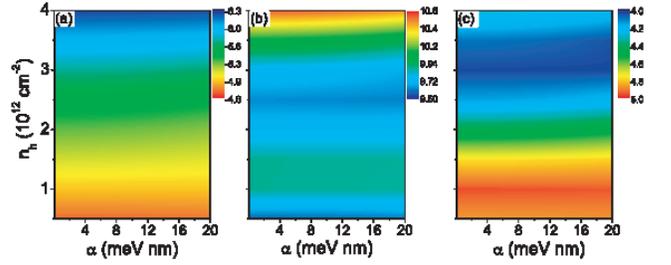}
\end{center}
\caption
{(Color online) (a) conventional spin Hall conductance component $\sigma
_{yx}^{s0}$, (b) spin torque dipole component $\sigma_{yx}^{s\tau}$, and (c)
the total spin Hall conductance $\sigma_{yx}^{s}$ (in unit of $e/4\pi$) as a
function of Rashba coefficient $\alpha$ and hole density $n_{h}$ (contour
plot).} \label{fig5}
\end{figure}%
Now we turn on the SIA Rashba term which is essential for understanding spin
transport in strong spin-orbit coupling regime in experimental quantum well
setup. Inclusion of $\alpha$-term in makes the analytic derivation of spin
Hall conductance very tedious. Instead of doing that, we numerically obtain
the eigenstates and eigen-energies of Hamiltonian (\ref{E24}), and substitute
them in Eq. (\ref{E17}) and (\ref{E19}). After a $k$-integration with fermi
wavevector pinned by the given hole density at zero temperature [see Fig. 1(b)
for the relationship between HH fermi wavevectors and $\alpha$], the spin Hall
conductance is therefore steadily calculated in a wide range of system
parameters. Figures 3(a)-(c) summarize the behavior of the conventional, the
spin-torque-dipole contributed, and the total spin Hall conductance,
respectively, as a function of Rashba coefficient $\alpha$ and hole density
$n_{h}$ (contour plot). Compared to the results without consideration of
Rashba spin splitting (Fig. 2), the prominent new features occurred in Figs. 3
are: (i) The conventional term $\sigma_{yx}^{s0}$ and the spin torque dipole
term $\sigma_{yx}^{s\tau}$ jump to much larger values upon switch on of Rashba
spin splitting. This jump is caused by the presence of coupling between the
two HH bands, which is absent without considering the SIA term; (ii) The
amplitude of $\sigma_{yx}^{s\tau}$ becomes larger than that of conventional
one $\sigma_{yx}^{s0}$ with a sign difference. In fact, the typical value of
$\sigma_{yx}^{s0}$ in a wide range in ($\alpha,n_{h}$) parameter space is
about ${\small \sim}\frac{-5e}{4\pi}$, while $\sigma_{yx}^{s\tau}$ is
typically of ${\small \sim}\frac{10e}{4\pi}$, resulting in $\sigma_{yx}%
^{s}{\small \sim}\frac{5e}{4\pi}$. Thus overall the conserved spin Hall
conductance changes a sign compared to the previous calculations based on
conventional definition; (iii) The stripe in Figs. 3 reveals that the
amplitude of intrinsic spin Hall conductance for 2DHG is insensitive to the
amplitude of Rashba coefficient $\alpha$, which implies a universal character.
The previous studies\cite{Sinova} on a $k$-linear Rashba 2DEG system within
the conventional definition of spin current have revealed a universal spin
Hall conductance ($\sigma_{yx}^{s0}$) of $e/8\pi$. We have also calculated the
conserved spin Hall conductance $\sigma_{yx}^{s}$ with the same model and
found it to be $-e/8\pi$. Together with the above results for 2DHG, one can
see the key role played by the spin torque dipole term, which tends to
overwhelm the conventional spin conductance by the opposite contribution.

Since the LH bands are usually unoccupied by the holes in the experiments, and
one can expect that in the weak Rshba spin splitting, the major contribution
to the spin Hall conductance comes from the coupling between the two HH bands.
In this case we can approximate the four-band Hamiltonian (\ref{E24}) by an
effective two-band one. After a standard adiabatic elimination procedure, we
obtain an effective two-band Hamiltonian $\bar{H}$ for heavy holes (in the
strong confinement limit $k^{2}<\langle k_{z}^{2}\rangle$),%
\begin{equation}
\bar{H}=\frac{\langle k_{z}^{2}\rangle}{2m_{HH}}+\frac{k^{2}}{2m^{\ast}%
}+i\beta(k_{-}^{3}\sigma_{+}-k_{+}^{3}\sigma_{-})\tag{30}\label{E30}%
\end{equation}
where $\beta=\frac{3}{8}\frac{\alpha(\langle k_{z}^{2}\rangle-4m_{LH}%
^{2}\alpha^{2})}{\langle k_{z}^{2}\rangle^{2}}$, $m_{HH,LH}=m/(\gamma_{1}%
\mp2\gamma_{2})$, $m^{\ast}=2m_{HH}m_{LH}/(m_{HH}+m_{LH})$ is the reduced mass
which is due to the proper account for the coupling between HH and LH bands,
and $\sigma_{\pm}=\sigma_{x}\pm i\sigma_{y}$ are Pauli matrices. To reflect
the angular momentum quantum numbers of the heavy holes, the conserved spin
Hall current operator is defined by $\hat{\boldsymbol{\mathcal{J}}}_{s}%
=\frac{3}{2}\frac{\mathrm{d}(\hat{\mathbf{r}}\hat{s}_{z})}{\mathrm{d}t}$. Thus
after an adiabatic projection onto the two HH bands, the Hamiltonian
(\ref{E24}) is reduced to an effectice $k$-cubic Rashba model. In this case,
we can derive an analytic expression for $\sigma_{yx}^{s}$ and its two
components $\sigma_{yx}^{s0}$ and $\sigma_{yx}^{s\tau}$. The results at zero
temperature are given by
\begin{align}
\sigma_{yx}^{s0} &  =-\frac{9e}{32\pi}\frac{\gamma_{1}}{m\beta}(\frac{1}%
{k_{f}^{+}}-\frac{1}{k_{f}^{-}})\nonumber\\
\sigma_{yx}^{s\tau} &  =\frac{27e}{32\pi}\frac{\gamma_{1}}{m\beta}(\frac
{1}{k_{f}^{+}}-\frac{1}{k_{f}^{-}})-\frac{9e}{8\pi}\tag{31}\label{E31}\\
\sigma_{yx}^{s} &  =\frac{9e}{16\pi}\frac{\gamma_{1}}{m\beta}(\frac{1}%
{k_{f}^{+}}-\frac{1}{k_{f}^{-}})-\frac{9e}{8\pi}\nonumber
\end{align}
where $k_{f}^{+}$ and $k_{f}^{-}$ are fermi wavevectors of the two HH bands.
According to the experimentally accessible hole density $n_{h}$, these two
fermi wave vectors can be expressed as\cite{Loss}
\begin{align*}
k_{f}^{+,-} &  =\left[  -\frac{1}{2}\left(  \frac{\gamma}{2\beta}\right)
^{2}\left(  1-\sqrt{1-\left(  \frac{2\beta}{\gamma}\right)  ^{2}4\pi n_{h}%
}\right)  +3\pi n_{h}\right]  ^{1/2}\\
&  \mp\frac{1}{2}\frac{\gamma}{2\beta}\left(  1-\sqrt{1-\left(  \frac{2\beta
}{\gamma}\right)  ^{2}4\pi n_{h}}\right)  .
\end{align*}
where $\gamma=\hbar^{2}/2m^{\ast}$. In the limit $n_{h}<<(1/4\pi
)(\gamma/2\beta)^{2}$, these fermi wavevectors have more simple forms:
$k_{f}^{+,-}\approx\sqrt{2\pi n_{h}}\mp(2\beta/\gamma)\pi n_{h}$. In this case
one gets $\sigma_{yx}^{s0}=-\frac{9e}{8\pi}$, $\sigma_{yx}^{s\tau}=\frac
{9e}{4\pi}$, thus $\sigma_{yx}^{s}=\frac{9e}{8\pi}$. In general $\sigma
_{yx}^{s}$ depends on the spin-orbit coupling and the fermi energy which is
pinned by the hole density. For further illustration, we show in Fig. 4%
\begin{figure}[tbp]
\begin{center}
\includegraphics[width=1.0\linewidth]{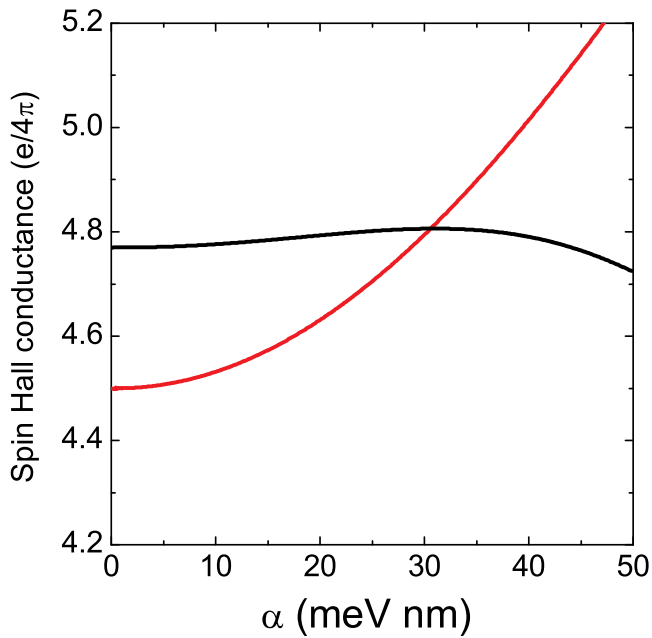}
\end{center}
\caption{(Color online) conserved spin Hall conductivity $\sigma_{yx}%
^{s}$ as a function of Rashba
coefficients $\alpha$ for $n_{h}=1.5\times10^{12}$ cm$^{-2}$. The black line
is the exact four-band result while the red line gives the approximate
two-band result for the heavy holes.} \label{fig4}
\end{figure}
the numerical results of conseved spin Hall conductance $\sigma_{yx}^{s}$ as a
function of Rashba coefficient $\alpha$ for the four-band (black curve) and
the approximate two-band (red curve) Hamiltonians. Note that the coefficient
$\beta$ in Eq. (\ref{E30}) is obtained from $\alpha$ by the equation below Eq.
(\ref{E30}). It reveals in Fig. 4 that the results from the two models agree
reasonably in the experimentally relevant variation of spin-orbit coupling
coefficient. When the value of spin splitting $\alpha k_{f}$ ($k_{f}$
characterizes fermi wavevector) is large enough to be comparable with the
confinement characteristic energy $\langle k_{z}^{2}\rangle/2m$, then the
deviation of the two-band approximation from the four-band calculation becomes
obvious, which is featured in Fig. 4 by the fact that while the two-band
$\sigma_{yx}^{s}$ increases monotonically as a function of $\alpha^{2}$ (which
can also be seen from Eq. (\ref{E31}), the four-band spin Hall conductance
displays a much weaker dependence of Rashba coefficient. Therefore we arrive
at a conclusion that the neglect of HH-LH coupling is no longer valid for
strong spin splitting and a more exact multi-band simulation is necessary for
the general 2DHG systems.

\section{Summary}

In this article we have studied spin transport features in general spin-orbit
coupled systems. Due to the spin-orbit coupling, the spin is not conserved by
the occurrence of spin torque during its flow through the sample. Therefore,
how to describe the spin transport within an intuitive drive-then-flow picture
becomes an important issue. The key point clarified in this paper is as
follows: (i) We have shown within the linear response formalism the necessity
to include the spin torque dipole term in the expression of the spin current.
The real advantage of our new definition of spin current lies in the fact that
it provides a satisfactory description of spin transport in the bulk. With our
new spin current, one can now use the spin continuity equation (\ref{E23}) to
discuss spin accumulation in the bulk, e.g., by generating a non-uniform
electric field or spatially modulating the spin Hall conductivity. Our new
spin current vanishes in Anderson insulators either in equilibrium or in a
weak electric field, which enables us to predict zero spin accumulation in
such systems. More importantly, it possesses a conjugate force (spin force),
so that spin transport can be fitted into the standard formalism of near
equilibrium transport. The conventional spin current does not have a conjugate
force, so it makes no sense even to talk about energy dissipation from that
current. The existence of a conjugate force is crucial for the establishment
of Onsager relations between spin transport and other transport phenomena, and
its measurement will be important to thermodynamic and electric determination
of the spin current. (ii) A general Kubo formula for the conserved spin
transport coefficients, consisting of the conventional and spin-torque-dipole
contrbutions, has been derived, which makes the practical bulk calculation to
be feasible.

Based on the Kubo formula for the conserved spin current, we have analyzed in
detail the spin Hall effect in 2DHG system with the parameters chosen to be
relevant to recent experimental measurement on GaAs quantum well and modeled
by the Hamiltonian consisting of a Luttinger spin-orbit coupling term and a
SIA Rashba term. In the absence of Rashba spin splitting, the two HH (and LH)
bands are degenerate and the non-zero spin Hall conductance only comes from
HH-LH transition. In this case, we have derived an analytic expression for the
conserved spin Hall conductance $\sigma_{yx}^{s}$ and its two components, the
conventional one $\sigma_{yx}^{s0}$ and the spin torque dipole correction
$\sigma_{yx}^{s\tau}$. These two components have been shown to compete each
other, which is verified by a difference of sign between them. In the case
that only the Luttinger term is taken into account, it has been found that the
amplitude of $\sigma_{yx}^{s\tau}$ is much small than that of $\sigma
_{yx}^{s0}$. In fact, the value of $\sigma_{yx}^{s0}$ is typically of
$-\frac{0.8e}{4\pi}$ while $\sigma_{yx}^{s\tau}$ is about $\frac{0.2e}{4\pi}$.
Thus in this case, the spin torque dipole correction to the spin Hall
conductivity is relatively small. When the SIA Rashba spin splitting is turned
on, we have found that there occurs a jump in amplitude for both $\sigma
_{yx}^{s0}$ and $\sigma_{yx}^{s\tau}$. $\sigma_{yx}^{s0}$ changes now to take
a typical value of $-\frac{5e}{4\pi}$ while $\sigma_{yx}^{s\tau}$ jumps to a
characteristic value of $\frac{10e}{4\pi}$. Therefore, the presence of Rashba
term not only stirs up a large enhancement of the total spin Hall conductivity
and its two components, but also changes a fundamental sign for $\sigma
_{yx}^{s}$ because the spin torque dipole correction now overwhelms the
conventional contribution. This jump and a change of sign for $\sigma_{yx}%
^{s}$ comes from the HH-HH coupling, and therefore can be modeled by an
effective two-band Hamiltonian which has been done in this paper by
adiabatically eliminating the LH states and reducing the system to a $k$-cubic
Rashba model with a dressed spin-orbit coupling coefficient [$\beta$ in Eq.
(\ref{E30})]. By a detailed comparison, furthermore, we have shown that this
two-band $\sigma_{yx}^{s}$ displays a quadric dependence of original Rashba
coefficient $\alpha$, while the exact four-band treatment shows a somewhat
universal character for $\sigma_{yx}^{s}$. Thus the neglect of HH-LH coupling
is no longer valid for strong spin splitting and a more exact multi-band
simulation is necessary for the general 2DHG systems.

\begin{acknowledgments}
PZ was supported by CNSF under Grant No. 10544004 and 10604010. JS was
supported by the \textquotedblleft BaiRen\textquotedblright\ program of the
Chinese Academy of Sciences. QN and DX were supported by DOE
(DE-FG03-02ER45958) and the Welch Foundation.\ 
\end{acknowledgments}

\end{document}